\documentclass[epjCONF]{svjour}
\usepackage{graphics}
\usepackage[varg]{txfonts} 
\usepackage[latin1]{inputenc}
\session-title{Conference Title, to be filled}
\begin{document}
\title{A search for binary candidates among the fundamental mode RR Lyrae stars observed by \textit{Kepler}}
\author{Elisabeth Guggenberger\inst{1,2}\fnmsep\thanks{\email{guggenberger@mps.mpg.de}} \and Jakob Steixner\inst{3} }
\institute{Max Planck Institute for Solar System Research, G\"ottingen, Germany \and Stellar Astrophysics Centre, Department of Physics and Astronomy, Aarhus University, Denmark \and Department of German Studies, University of Vienna}
\abstract{
Although roughly half of all stars are considered to be part of binary or multiple systems, there are only two confirmed cases of RR Lyrae pulsators with companions. One of them is TU Uma \cite{wade} - a classical RR Lyrae star in a very eccentric orbit - and the other is OGLE-BLG-RRLYR-02792 \cite{pietr}. 
Considering the wealth of well-studied RR Lyrae stars, this number is astoundingly low. 
Having more RR Lyrae stars in binary systems at hand would be extremely valuable to get independent measurements of the masses. The data from the \textit{Kepler} mission with their unprecedented precision and the long time span of about four years offer a unique possibility to systematically search for the signatures of binarity in RR Lyrae stars. Using the pulsation as a clock, we studied the variations in the timing of maximum light to hunt for possible binary systems in the sample. 
} 
\maketitle

\section{Methods and data}
\label{data}
For our search we made use of the full time span of data provided by the \textit{Kepler} mission, i.e., about 4 years (Q0-Q17). All the stars in our sample have been observed in long cadence (LC, $\sim$29 min sampling) during the whole operational time of the mission. Short cadence (SC, $\sim$1 min sampling) data are available at least during one quarter for each star.

About half of all RR Lyrae stars pulsating in the radial fundamental mode (RRab stars) are known to be modulated by the so-called Blazhko effect. This phenomenon can lead to periodic phase or period changes which would be hard to disentangle from any changes due to binarity. Even though this would in principle be possible, we here restricted our search to the 18 non-Blazhko stars in the sample. A detailed analysis of the Q0-Q5 data of these stars has been published by \cite{nemec}, but no search for a light time effect or period fluctuations had been performed.

For our study we used the classical O-C technique, where O stands for ``observed'', and C for ``calculated''. The time of maximum light is measured for each pulsation cycle, yielding the "O" value. On the other hand, an expected time of maximum, "C", is predicted assuming a constant pulsation period. The difference between the two values is of great diagnostic value, as period changes caused by binarity, continuous period changes due to evolution, and abrupt changes all cause different patterns.

To measure the exact time of maximum light from the light curves, a polynomial fit was computed for each pulsation cycle. As most of the data were available in LC only, and RRab stars have sharp maxima, a fit with free parameters showed large scatter. It turned out to be very sensitive to the location of the data points with respect to the light maximum. Fitting a template (i.e., a polynomial with fixed parameters) is a more robust method to accurately determine the time of maximum light even with sparse sampling. We created templates for each star based on the SC data. These templates were then cross-correlated with the LC light curves. The resulting scatter in the O-C diagram is about 1 minute. Using phase diagrams of three periods decreased the scatter to below one minute in most of the cases. 

\section{Results}
\label{results}
Variations in the O-C diagrams were detected for several stars in the sample. Three stars (KIC 07176080, KIC 07030715 and KIC 09591503, see Fig. 1) show an O-C variation that could be interpreted as periodic and which might be indicative of binarity. However, even though only bona-fide non-modulated stars were used for this study, it is possible that an intrinsic stellar modulation is present that manifests itself as pure phase modulation. As the mechanism behind the Blazhko effect is still unknown it cannot be excluded that purely phase-modulated stars exist and mimic the O-C variations of binary stars. 

Additionally, RR Lyrae stars can show irregular fluctuations and sudden jumps in their periods. These phenomena could possibly also be mistaken for periodic, when observed only over a limited time span.
In our sample we find one star with a sudden period jump (KIC 11802860) and several cases of variation that seem irregular (for example KIC 8344381).  \cite{swei} suggested that mixing events might be held responsible for such period jumps in RR Lyrae stars, as well as for period variations that are too fast for evolution. \cite{leborgne} have also reported erratic O-C variations for some of their targets in their study on evolutionary period changes in RR Lyrae stars. 

Several stars in our sample show the typical horizontal flat line which indicates that the period has remained constant (for example KIC 05299596 and KIC 6100702). The star is neither a binary nor modulated (or the variation is too small to be detected). 

We note that the amplitude of the O-C fluctuations are very small (a few minutes) while the periods of the candidates are long (in the order of years), indicating that if they are indeed binaries, the inclination would be very low. Any attempt to follow up the targets spectroscopically would therefore be very challenging, especially given the faintness of the stars (13.3 - 17.4 mag).%

\begin{figure}
\resizebox{\columnwidth}{!}{ 
\includegraphics{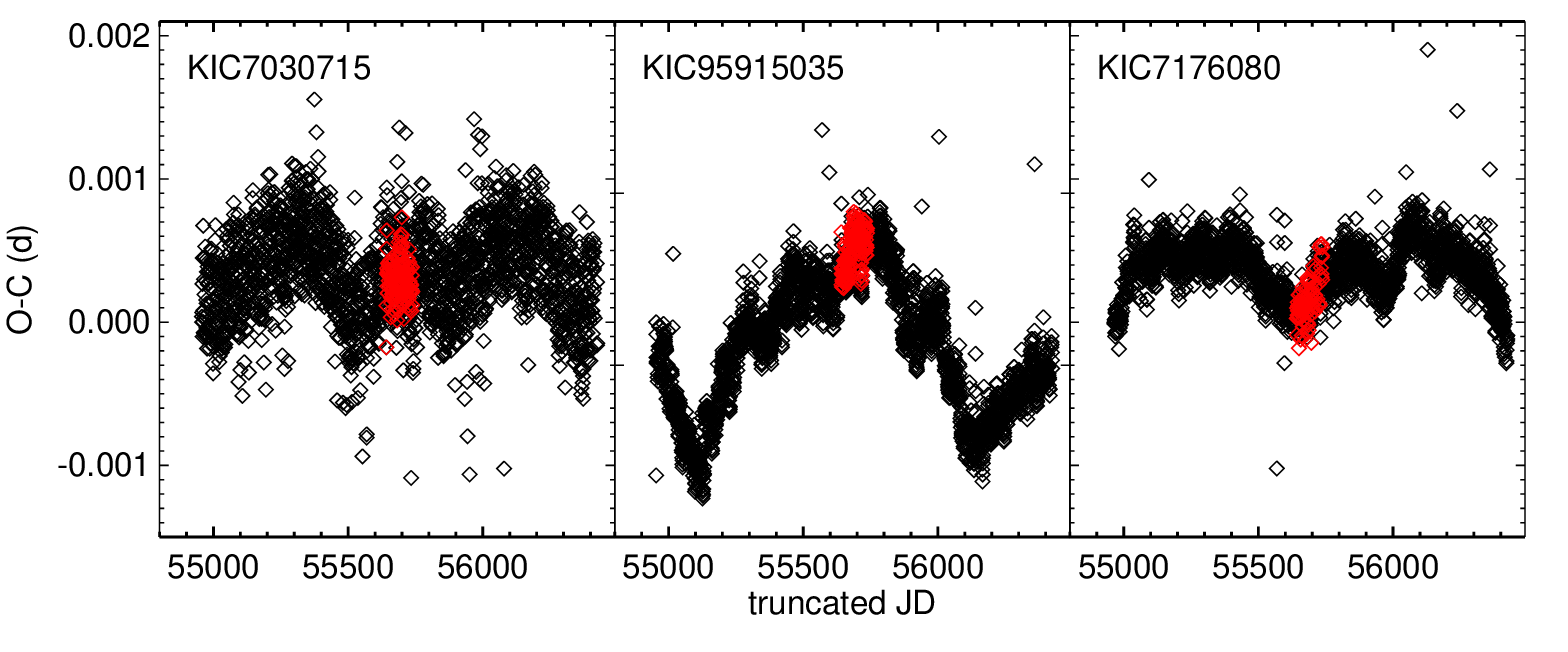} }
\caption{The three stars with a possible periodic O-C variation that might be caused by binarity. Long cadence data are shown in black, short cadence data in red.}
\label{fig:1}       
\end{figure}
\begin{acknowledgement}\textbf{Acknowledgements.}The research leading to the presented results has received funding from the European Research Council under the European Community's Seventh Framework Programme (FP7/2007-2013) / ERC grant agreement no 338251 (StellarAges).
\end{acknowledgement}

\end{document}